\documentclass[a4paper, oneside, twocolumn, notitlepage, 10pt]{extarticle_ecoc2015}
\usepackage{ecoc2015}
\usepackage{multirow}
\begin{document}
\selectlanguage{american}    


\title{400 Gbps Dual-polarisation Non-linear Frequency-division
Multiplexed Transmission with b-Modulation}%


\author{
    Xianhe~Yangzhang\textsuperscript{(1)(2)},  Vahid Aref\textsuperscript{(2)}, Son T. Le\textsuperscript{(2)}, Henning B\"ulow\textsuperscript{(2)},
    Polina~Bayvel\textsuperscript{(1)}
}

\maketitle                  


\begin{strip}
 \begin{author_descr}

   \textsuperscript{(1)} Optical Networks Group, Department of Electronic \& Electrical Engineering, UCL, UK
   \uline{x.yangzhang@ucl.ac.uk}

   \textsuperscript{(2)} Nokia-Bell-Labs, Stuttgart, Germany


 \end{author_descr}
\end{strip}

\setstretch{1.1}


\begin{strip}
  \begin{ecoc_abstract}
    We demonstrate, for the first time, a b-modulated dual-polarisation NFDM transmission in simulation, achieving a record net data rate of 400 Gbps (SE of 7.2 bit/s/Hz) over 960 km. The proposed scheme shows 1 dB Q-factor improvement over $q_c$-modulation scheme.
    
%
%
%
  \end{ecoc_abstract}
\end{strip}


\section{Introduction}
Nonlinear frequency-division multiplexing (NFDM) has been intensively investigated as a potential transmission scheme to overcome the performance limitation due to the Kerr effect in optical fibre. Under specific conditions, NFDM shows significantly weaker inter-channel interference in comparison to the conventional linear multiplexing schemes, e.g. wavelength-division multiplexing (WDM)\cite{yangzhang2018jlt}\cite{yousefi2016ecoc}.  However, despite many respectful endeavours \cite{Aref2018,le2018jlt}, only up to 2.3 bit/s/Hz \cite{le2018jlt} has been achieved in NFDM systems up to date due the two main reasons, namely i) most of NFDM systems up to date use only one polarization for data transmission and ii) severe signal-noise interaction \cite{Aref2018,le2018jlt}. 

To suppress the signal-noise interaction in NFDM transmissions, it is crucial to  find the ``correct" degrees of freedom offered by non-linear spectrum for data modulation. It has been shown (for single-polarization) that modulating information on the b-coefficient significantly improves the transmission performance compared to the conventional $q_c$-modulation scheme \cite{le2018ofc}\cite{wahls2017ecoc}.

In this work, for the first time, we developed a novel transform which enables $b$-modulation for dual-polarisation under the constraint $|b_1(\lambda)|^2+|b_2(\lambda)|^2<1$, which outperforms the conventional $q_c$-modulation \cite{Goossens17} scheme by over 1 dB. In addition, the b-modulated dual-polarisation transmission system is systematically optimised for increasing the data rate. As a resutl, we achieved in simulation a record net data rate of 400 Gbps with a SE of 7.2 bits/s/Hz over 12 spans of 80 km of standard single mode fibre (SSMF) with EDFAs. 


\section{The Optic-fibre Model}
The optic-fibre model of concern is a multi-span point-to-point dual-polarisation dispersion unmanaged fibre link with equally-spaced erbium-doped fibre amplifiers (EDFAs), which can be described by the Manakov equation
\begin{align*} 
\frac{\partial \vec{Q}}{\partial z}+\frac{\alpha}{2}\vec{Q}+\frac{j\beta_2}{2}\frac{\partial^2 \vec{Q}}{\partial t^2} -j\frac{8}{9}\gamma \vec{Q}\left\Vert\vec{Q}\right\Vert^2 = 0
\label{eq:edfa}
\end{align*}
where $j=\sqrt{-1}$ and $\vec{Q}= [q_x(t,z)~q_y(t,z)]$ is the complex envelope of the signal as a function of time $t$ and distance $z$ along the fibre, $D$, $\gamma$, and $\alpha$ are the dispersion, non-linear, and loss coefficients of the fibre; Tab.~\ref{tab:fibre para} lists all parameters. 
\begin{table}
   \centering
\caption{Fibre and system simulation parameters} \label{tab:fibre para}
\begin{tabular}{|c|c|c|}
  \hline
  $\nu$ & $193.44$ THz & centre carrier freq. \\
  $\alpha$ & $0.2~\rm{dB}~ \rm{km}^{-1}$ & fibre loss\\
  $\gamma$ & $1.3$ $\rm{(W\cdot km)}^{-1}$ & non-linearity para.\\
  $D$ & $16.89$ ps/(nm-km) & dispersion para.\\
  $W$ & 56 GHz &linear bandwidth \\
  $R_o$ & 8 & oversampling rate\\
  $\mathcal{L}_{\textnormal{sp}}$ & $80$ km & span length\\
  $N_{\textnormal{sp}}$ & 12 & number of spans\\
  NF & 5 dB & EDFA noise figure\\
  \hline
\end{tabular}
\end{table}%
\vspace{-0.2cm}
\section{$q_c$ and $b$-modulation}
\begin{figure}[htp!]
   \centering
    \includegraphics[width=0.4\textwidth]{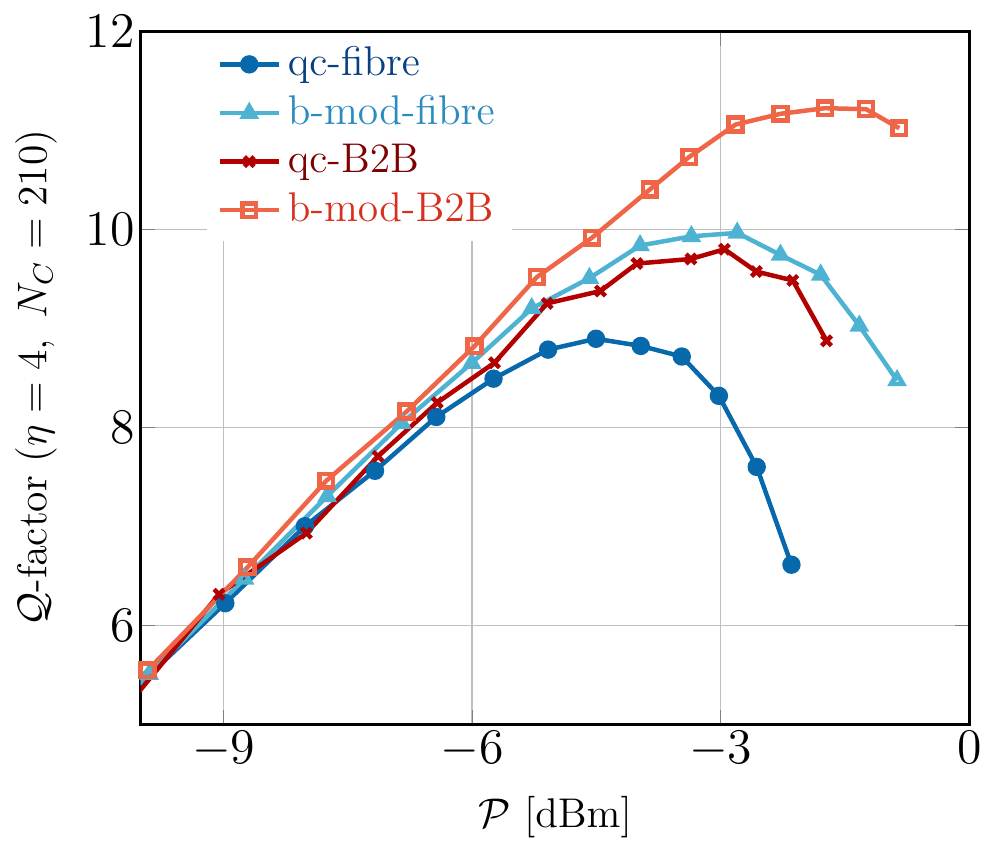}
    \caption{Comparison of $q_c$ and $b$ modulation with large guard interval $\eta=4$ in both back-to-back case and fibre transmission. The total additive noise powers are the same. }
    \label{fig:bqceta4}
\end{figure}

\begin{figure*}
   \centering
   \begin{tabular}{cc}
    \includegraphics[width=0.665\textwidth]{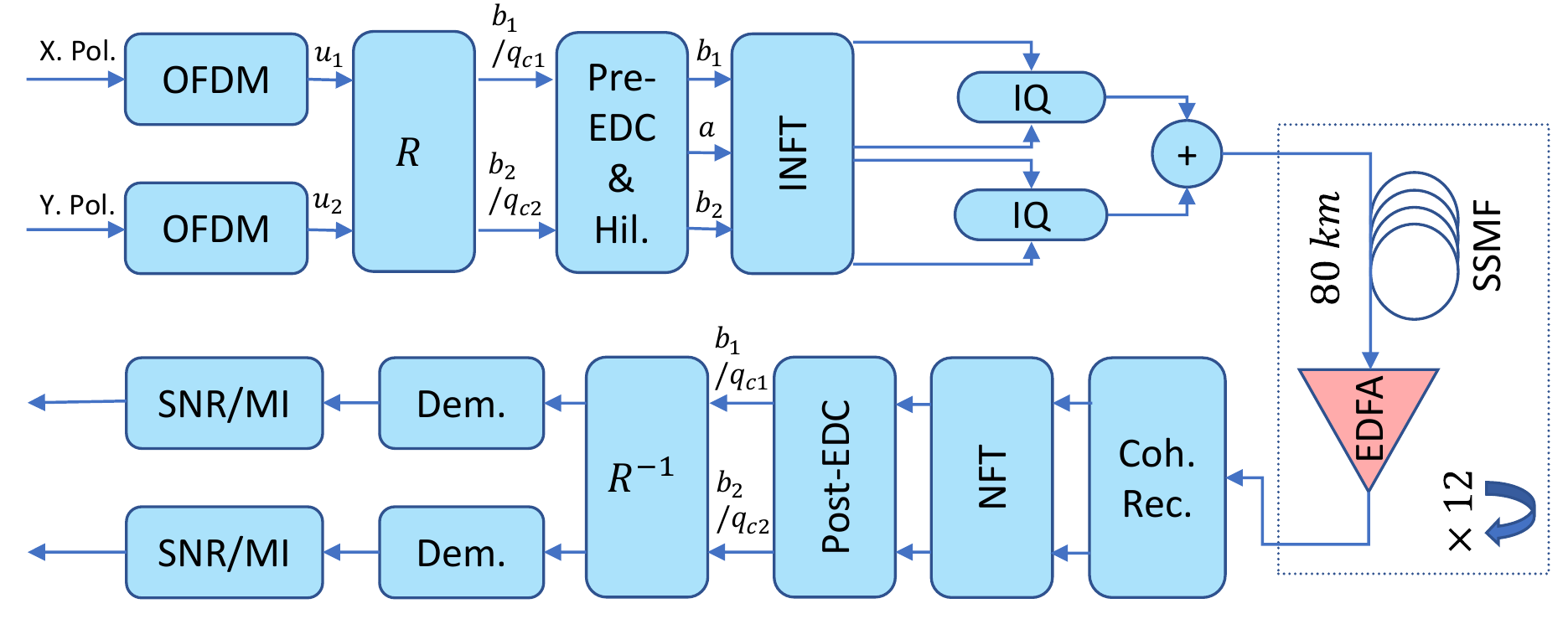} &
        \includegraphics[width=0.285\textwidth]{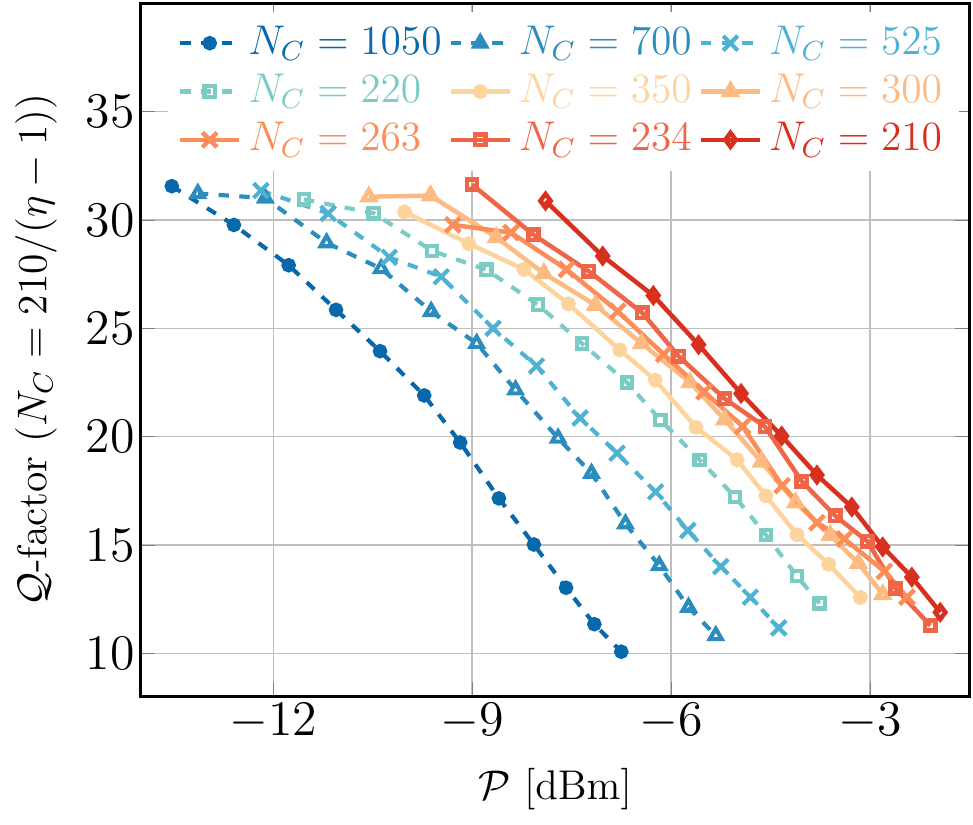}\\
        (a)&(b)\\
   \end{tabular}
    \caption{Simulation diagram (a) and its noiseless back-to-back characterisation (b). }
    \label{fig:balg}
\end{figure*}

We explain in this section $q_c$ and $b$-modulation, respectively. For each NFDM symbol, the standard OFDM is used to construct signal $u_i(\lambda)$
\begin{equation*} 
u_i(\lambda) = \sum_{k=-N_C/2}^{N_C/2-1}c^k_i \frac{\sin(\lambda T_0+k\pi)}{\lambda T_0+k\pi},~i \in \{1,2\},
\label{eq:orth}
\end{equation*}
where $k$ and $i$ is the sub-carrier and polarisation index,  $c_i^k$ is chosen from the  32-QAM constellation, $T_0 =N_C/W$, and $N_C$ is the number of sub-carriers. We define $\eta=(T_0+T_G)/T_0$, where $T_G$ is the guard interval. The discrete signal $\vec{u}_{i}=[u^1_{i},...,u^n_{i}]$ can be implemented by 
 \begin{equation*}
\begin{array}{l}
 \vec{d}_i=\textnormal{IDFT}\{f(\vec{c}_i,N_C(R_o-1))\},\\
 \vec{u}_{i}=\begin{cases}
\textnormal{DFT} \{f(\vec{d}_i,N_CR_o(\eta-1))\}, \eta\geq 2,\\
 \textnormal{DFT} \{f(\vec{d}_i,N_CR_o)\}, 1<\eta<2,\\
 \end{cases}
\end{array}
 \end{equation*}
where $R_o$ is the oversampling rate and $f(\vec{x},N)$ is defined as a function that adds $N$ zeros to vector $\vec{x}$ in the following way
\begin{equation*}
f(\vec{x},N) = [\underbrace{0,...,0}_{N/2}, x^1,...,x^n,\underbrace{0,...,0}_{N/2}].
\end{equation*}
 For $q_c$-modulation, the following transform denoted as $R$ in Fig.~\ref{fig:balg}(a) was performed for signals in both polarisations \cite{Goossens17}
 \begin{equation*}
  \begin{array}{rl}
q_{c1}(\lambda) &= \sqrt{e^{|u_1(\lambda)|^2}-1}\cdot e^{j\angle u_1(\lambda)},\\
q_{c2}(\lambda) &= \sqrt{e^{|u_2(\lambda)|^2}-1}\cdot e^{j\angle u_2(\lambda)}.\\
 \end{array}
 \end{equation*}
For $b$-modulation, the transform $R$ becomes
\begin{align*}
A & =\sqrt{ \frac{1-\exp(-|u_1(\lambda)|^2-|u_2(\lambda)|^2)}{|u_1(\lambda)|^2+|u_2(\lambda)|^2}},\\
b_1(\lambda) &= A\cdot u_1(\lambda),~b_2(\lambda)  = A\cdot u_2(\lambda),\\
a(\lambda) & = \exp(\mathbf{H}(\log(\sqrt{1-|b_1(\lambda)|^2-|b_2(\lambda)|^2}))),
\end{align*}
where $\mathbf{H}$ stands for the Hilbert transform, explained in \cite{wahls2017ecoc}. Subsequently, either $\{a(\lambda),b_1(\lambda),b_2(\lambda)\}$ or $\{q_{c1}(\lambda), q_{c2}(\lambda)\}$ will be passed through the inverse NFT to calculate the time domain signal. The transform $R$ for $b$ coefficient facilitates the signal modulation because $u(\lambda)$ ranges the entire complex plane, while $b_1(\lambda)$ and $b_2(\lambda)$ are constraint by $|b_1(\lambda)|^2+|b_2(\lambda)|^2<1$. 
Note that, for $\eta <2$, each burst will be truncated symmetrically to a signal of length $N_CR_0\eta$  for transmission and also recovered to length $2N_CR_0$ by adding zeros before NFT processing. At the receiver, the NFT and back rotation equalisation are performed.
We demonstrate first the advantage of $b$-modulation over $q_c$-modulation in a relatively ideal scenario, i.e., large guard interval. Fig.~\ref{fig:bqceta4} clearly shows that the non-linear processing penalty (back-to-back) of $b$-modulation is roughly 1.5 dB in Q-factor smaller than the penalty in $q_c$-modulation. When signals are transmitted through fibre, extra Q-factor degradation are observed due to the approximation error of path-averaged model in both cases. Note that, in all simulations, pre-compensation \cite{Tavakkolnia2016} was applied to reduce the temporal broadening caused by dispersion. The power $\mathcal{P}$  reported in this work is always per polarisation.

\section{Optimisation of $b$-modulated dual-polarisation NFT transmission}
Once the advantage of $b$-modulation is established, we optimise the total data rate over several system parameters providing the optical fibre channel specified by Tab.~\ref{tab:fibre para}. The available optimisation parameters are number of sub-carriers $N_C$, the ratio $\eta$ and launch power per polarisation $\mathcal{P}$. The optimisation process can be simplified by estimating the required guard interval\cite{le2018jlt}
\begin{equation*}\label{eq:gi}
T_G \approx \pi W\beta_2 L=3.75~\textnormal{ns}.
\end{equation*}
$T_G$ depends only on signal bandwidth and transmission distance and thus, should remain constant in this work. As defined earlier,
\begin{equation*}
\eta = (T_0+T_G)/T_0=1+WT_G/N_C.
\end{equation*}
For instance, if $\eta=2$, $N_C=210$. Any further reduction of $\eta$ should be achieved by increasing $N_C$, consequently the spectral efficiency (SE) loss due to guard interval will diminish. However, for large $N_C$, non-linear processing (NFT or INFT) penalty will become more severe, resulting in a Q-factor degradation. The non-linear processing penalty is attributed to two factors: 1) inaccuracy of algorithms for high-energy pulses, 2) energy-dependent signal-noise interaction.  
\begin{figure*}[htp]
\centering
	\begin{tabular}{ccc}
\includegraphics[width = 0.37\textwidth]{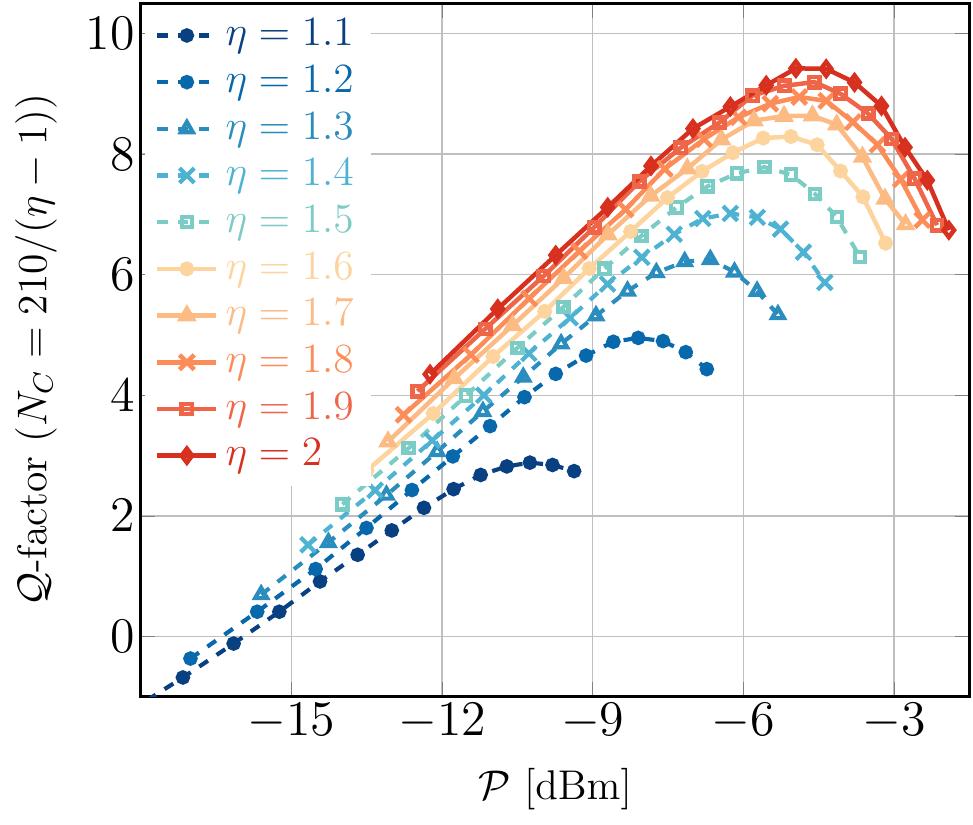}
	&
    \includegraphics[width = 0.37\textwidth]{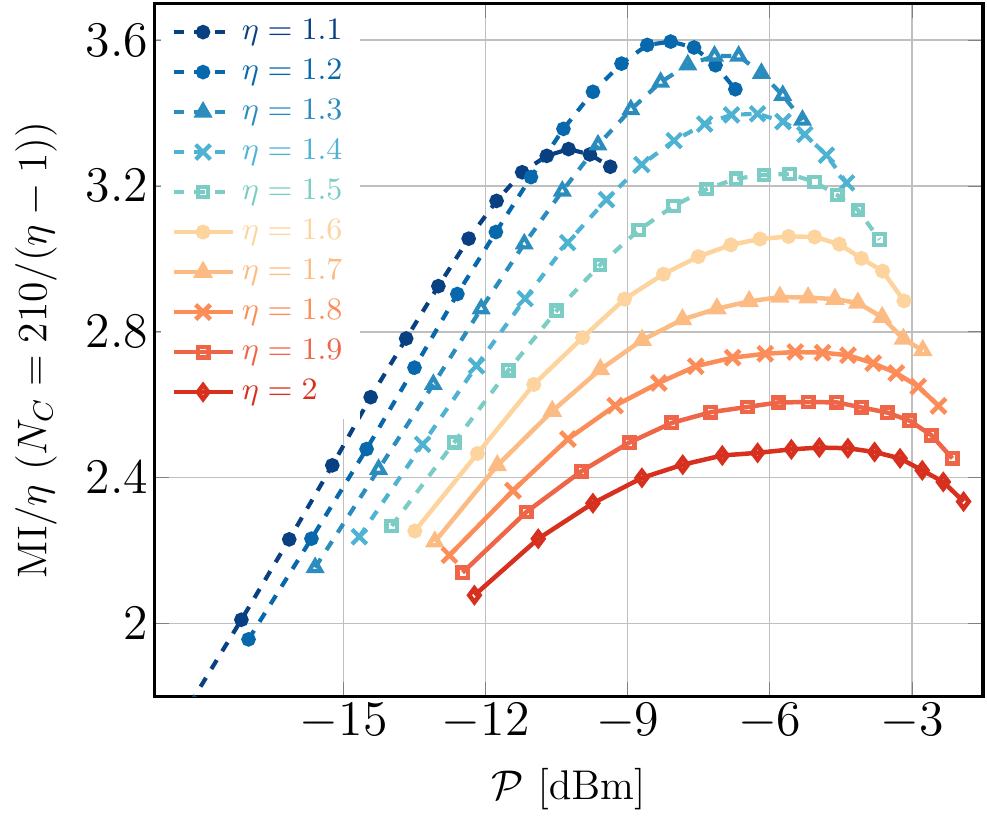}
    &
    \raisebox{0.8\height}{
      \begin{minipage}[t]{0.19\textwidth}
    \begin{tabular}{c}
        \includegraphics[width = 0.8\textwidth]{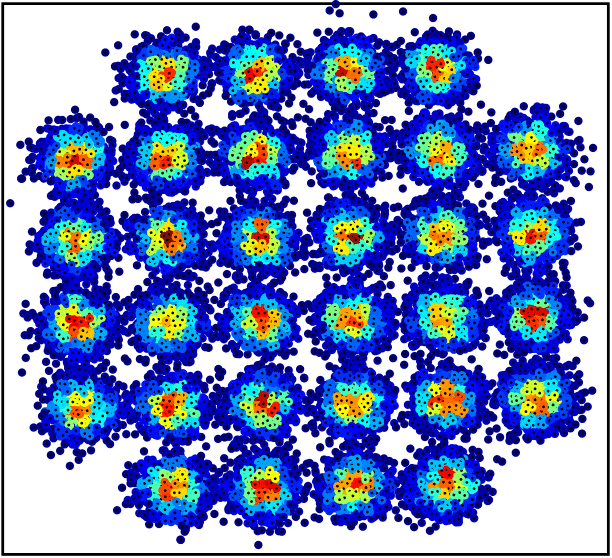}\\
        (c)\\
    \includegraphics[width = 0.8\textwidth]{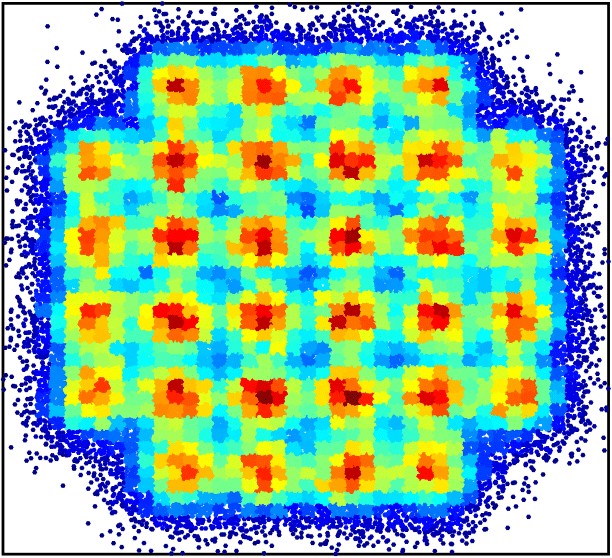}\\
    (d)
    \end{tabular}
  \end{minipage}}\\
  (a) & (b) &
	\end{tabular}	
	\caption{Further decrease of $\eta$ and its impact on Q-factor and MI$/\eta$ (SE). (c) is the received constellation at the point of maximum Q-factor in (a). (d) is the received constellation at the point of maximum MI/$\eta$ in (b).}
    \label{fig:etaopt}
\end{figure*}

Fig.~\ref{fig:etaopt} shows the further optimisation results with smaller $\eta$. It can be seen from Fig.~\ref{fig:etaopt}(a) that reducing $\eta$ will increase the non-linear processing penalty. In terms of spectral efficiency MI$/\eta$, the benefit of reducing $\eta$ is more significant than the information loss due to the increased non-linear processing penalty, until the threshold $\eta=1.2$ as shown in Fig.~\ref{fig:etaopt}(b).
Furthermore, we quantify the noise induced by algorithmic inaccuracy with simulations in the absence of additive noise. Fig.~\ref{fig:balg}(b) shows that the algorithmic inaccuracy becomes the primary limitation on the achievable mutual information rate at high power providing the same $R_o$ and $\eta$ and worsens with increasing $N_C$. The finding is also in consistence with the results in \cite{lima2015}\cite{Vaibhav2017}. The required spectral resolution increases as $|b_1(\lambda)|^2+|b_2(\lambda)|^2$ approaches $1$. Finally, by choosing $\eta=1.2$, at power of $-8$ dBm per polarisation, we achieved a spectral efficiency of $3.6$ bits/s/Hz per polarisation, resulting $400$ Gbps net data rate, while the gross data rate is $560$ Gbps. Should the algorithmic accuracy be improved, higher data rate or SE is expected .


\section{Conclusions}
In conclusion, our contribution in this work is threefold: 1) we developed a transform that enables polarisation-multiplexed $b$-modulation under the constraint $|b_1(\lambda)|^2+|b_2(\lambda)|^2<1$, 2) we demonstrated that,  when using the continuous non-linear spectrum, modulating $b$ coefficient instead of $q_c$ can provide a significant performance improvement in polarisation-multiplexed NFDM system, 3) We optimised the system for higher data rate and achieved a record high net data rate of 400 Gbps with a SE of 7.2 bits/s/Hz. Higher data rate can be expected once we improve the algorithmic accuracy.

\section{Acknowledgements}

The work is within the COIN project, financed by the European Commission grant
676448, under the call H2020-MSCA-ITN-2015.



\vspace{-4mm}

\end{document}